\documentclass[twoside,a4paper,11pt]{sea10}
\usepackage{graphicx}
\usepackage{hyperref}
\usepackage{movie15}
\newcommand\arcsec{\mbox{$^{\prime\prime}$}}
\newcommand\arcmin{\mbox{$^\prime$}}
\begin{document}
\pagenumbering{arabic}
\pagestyle{myheadings}
\thispagestyle{empty}
{\flushleft\includegraphics[width=\textwidth,bb=58 650 590 680]{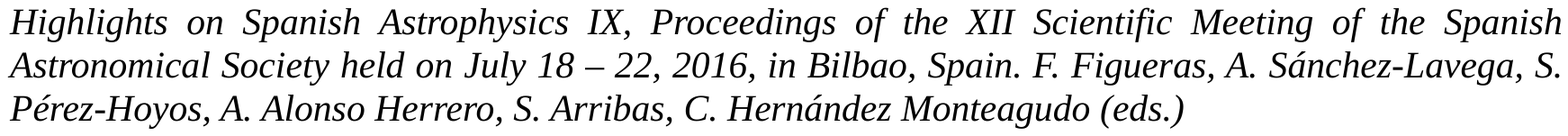}}
\vspace*{0.2cm}
\begin{flushleft}
{\bf {\LARGE
%
Progress towards a universal family of UV-IR extinction laws
%
}\\
\vspace*{1cm}
%
Jes\'us Ma{\'\i}z Apell\'aniz$^{1}$,
Emilio Trigueros P\'aez$^{2}$, 
Azalee K. Bostroem$^{3}$, 
Rodolfo H. Barb\'a$^{4}$, 
and 
Christopher J. Evans$^{5}$
%
}\\
\vspace*{0.5cm}
%
$^{1}$
Centro de Astrobiolog\'{\i}a, CSIC-INTA, Madrid, Spain\\
$^{2}$
Universidad Complutense, Madrid, Spain\\
$^{3}$
University of California Davis, California, U.S.A.\\
$^{4}$
Universidad de La Serena, La Serena, Chile\\
$^{5}$
U.K. Astronomy Technology Centre, ROE, Edinburgh, U.K.
%
\end{flushleft}
%
\markboth{
UV-IR extinction laws
}{ 
%
Ma{\'\i}z Apell\'aniz et al.
%
}
\thispagestyle{empty}
\vspace*{0.4cm}
\begin{minipage}[l]{0.09\textwidth}
\ 
\end{minipage}
\begin{minipage}[r]{0.9\textwidth}
\vspace{1cm}
\section*{Abstract}{\small
%
We present our progress on the study of extinction laws along three diferent lines. [a] We compare how well different families
of extinction laws fit existing photometric data for Galactic sightlines and we find that the 
\href{http://adsabs.harvard.edu/abs/2014A&A...564A..63M}{Ma{\'{\i}}z Apell{\'a}niz et al. (2014)} family provides better results than
those of \href{http://adsabs.harvard.edu/abs/1989ApJ...345..245C}{Cardelli et al. (1989)} or 
\href{http://adsabs.harvard.edu/abs/1999PASP..111...63F}{Fitzpatrick (1999)}. [b] We describe the HST/STIS spectrophotometry in 
the 1700-10\,200~\AA\ range that we are obtaining for several tens of sightlines in 30 Doradus with the purpose of deriving an
improved wavelength-detailed family of extinction laws. [c] We present the study we are conducting on the behavior of the extinction
law in the infrared by combining 2MASS and WISE photometry with Spitzer and ISO spectrophotometry.
%
\normalsize}
\end{minipage}
%
%

\section{Don't we know everything about extinction laws already?}

$\,\!$\indent The answer to that question is no, not really. Here are some of the pending issues regarding extinction laws:

\begin{itemize}
 \item In the mid-infrared (MIR): how do the water, aliphates, and silicates features vary among sightlines?
 \item In the near-infrared (NIR): what is the slope of the power law that is commonly used to characterize the extinction law? Is it universal?
 \item In the optical range: what is the functional form of the extinction law? How do you include the diffuse interstellar bands (DIBs)? 
       Can we find sightlines to measure the extinction law from 3000~\AA\ to 30~$\mu$m continuously?
 \item Are the IR/optical and the ultraviolet (UV) extinction laws correlated or not?
\end{itemize}

Some of the problems above are related to the lack of data e.g. very few UV sightlines have been measured at low metallicity. Others are due to 
the existence of published works with calibration issues or with outdated methods. Once we solve those questions we can attack with confidence more
profound ones such as how the extinction law varies as a function of the type of environment. Considering that extinction affects a large number of astronomical
measurements, it is important we undersatnd how to correct for it.

\section{What are we doing about it?}

$\,\!$\indent Here we describe our work on the field of extinction. We have already published the following:

\begin{itemize}
 \item A new family of optical/NIR extinction laws (\href{http://adsabs.harvard.edu/abs/2014A&A...564A..63M}{Ma{\'\i}z Apell\'aniz et al. 2014a}) based on 
       modern 30 Doradus data.
 \item New UV sightlines in the SMC that are different from previous ones 
       (\href{http://adsabs.harvard.edu/abs/2012A&A...541A..54M}{Ma{\'\i}z Apell\'aniz \& Rubio 2012}).
 \item A preliminary analysis of the IR extinction (\href{http://adsabs.harvard.edu/abs/2015hsa8.conf..402M}{Ma{\'\i}z Apell\'aniz 2015a}).
 \item The relationship between the optical DIBs and extinction (\href{http://adsabs.harvard.edu/abs/2014IAUS..297..117M}{Ma{\'\i}z Apell\'aniz et al. 2014b},
       \href{http://adsabs.harvard.edu/abs/2015MmSAI..86..553M}{Ma{\'\i}z Apell\'aniz 2015b}).
 \item Studies of specific sightlines (\href{http://adsabs.harvard.edu/abs/2006MNRAS.366..739A}{Arias et al. 2006}, 
       \href{http://adsabs.harvard.edu/abs/2015A&A...579A.108M}{Ma{\'\i}z Apell\'aniz et al. 2015a}, 
       \href{http://adsabs.harvard.edu/abs/2015A&A...583A.132M}{2015b}, 
       \href{http://adsabs.harvard.edu/abs/2015hsa8.conf..604M}{2015c}).
 \item Different calibration issues (\href{http://adsabs.harvard.edu/abs/2004PASP..116..859M}{Ma{\'\i}z Apell\'aniz 2004}, 
       \href{http://adsabs.harvard.edu/abs/2005PASP..117..615M}{2005a}, 
       \href{http://adsabs.harvard.edu/abs/2006AJ....131.1184M}{2006}, 
       \href{http://adsabs.harvard.edu/abs/2007ASPC..364..227M}{2007}, 
       \href{http://adsabs.harvard.edu/abs/2013hsa7.conf..583M}{2013a}, 
       \href{http://adsabs.harvard.edu/abs/2013hsa7.conf..657M}{2013b}).
\end{itemize}

We currently have three different open lines and in this contribution we discuss their status. They are:

\begin{itemize}
 \item Comparing different families of extinction laws in the Galaxy.
 \item STIS 1700-10\,200~\AA\ spectrophotometry of 30 Doradus OB stars.
 \item A study of the IR extinction with photometry and spectrophotometry.
\end{itemize}

\begin{figure}
\centerline{\includegraphics*[width=0.55\linewidth]{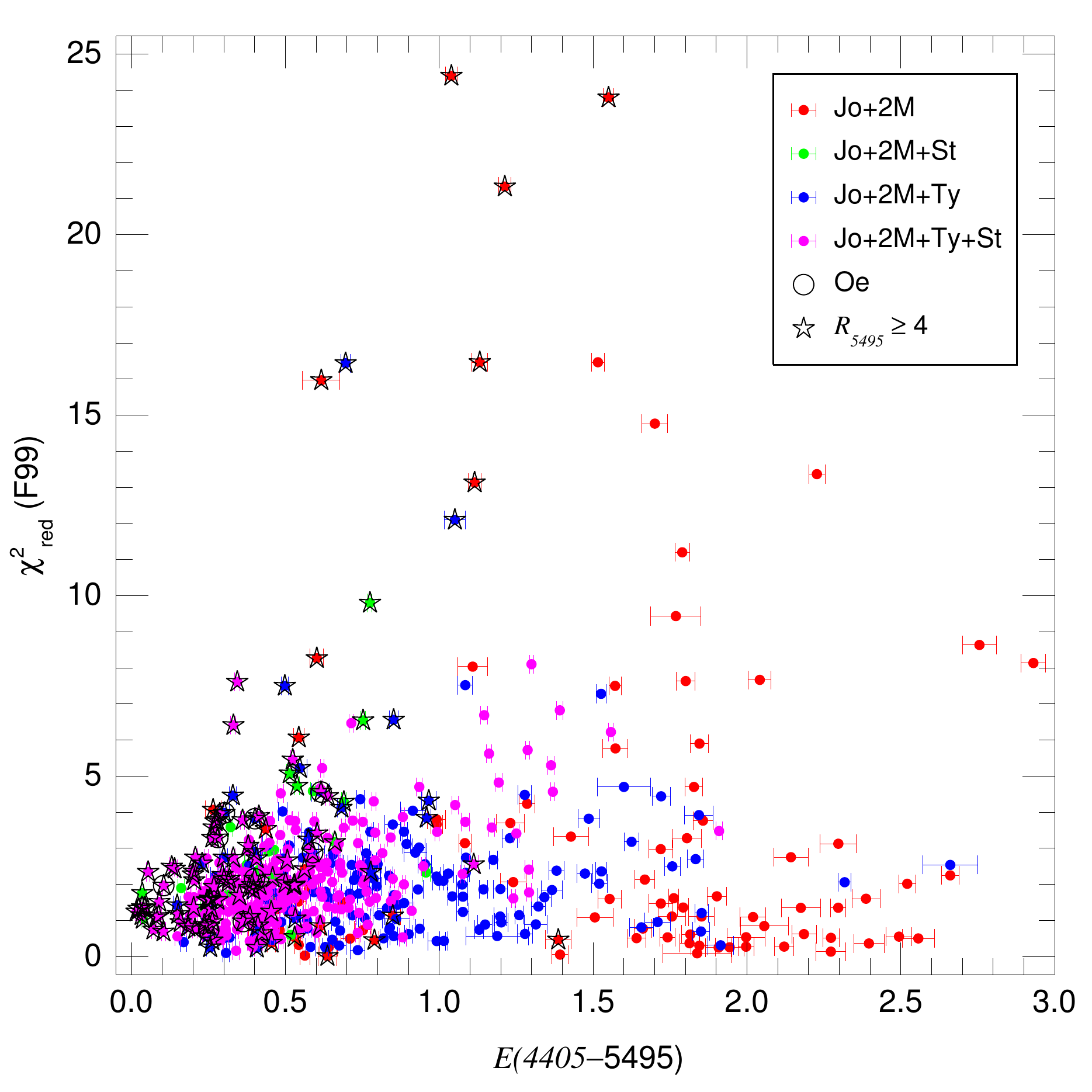}\,
            \includegraphics*[width=0.55\linewidth]{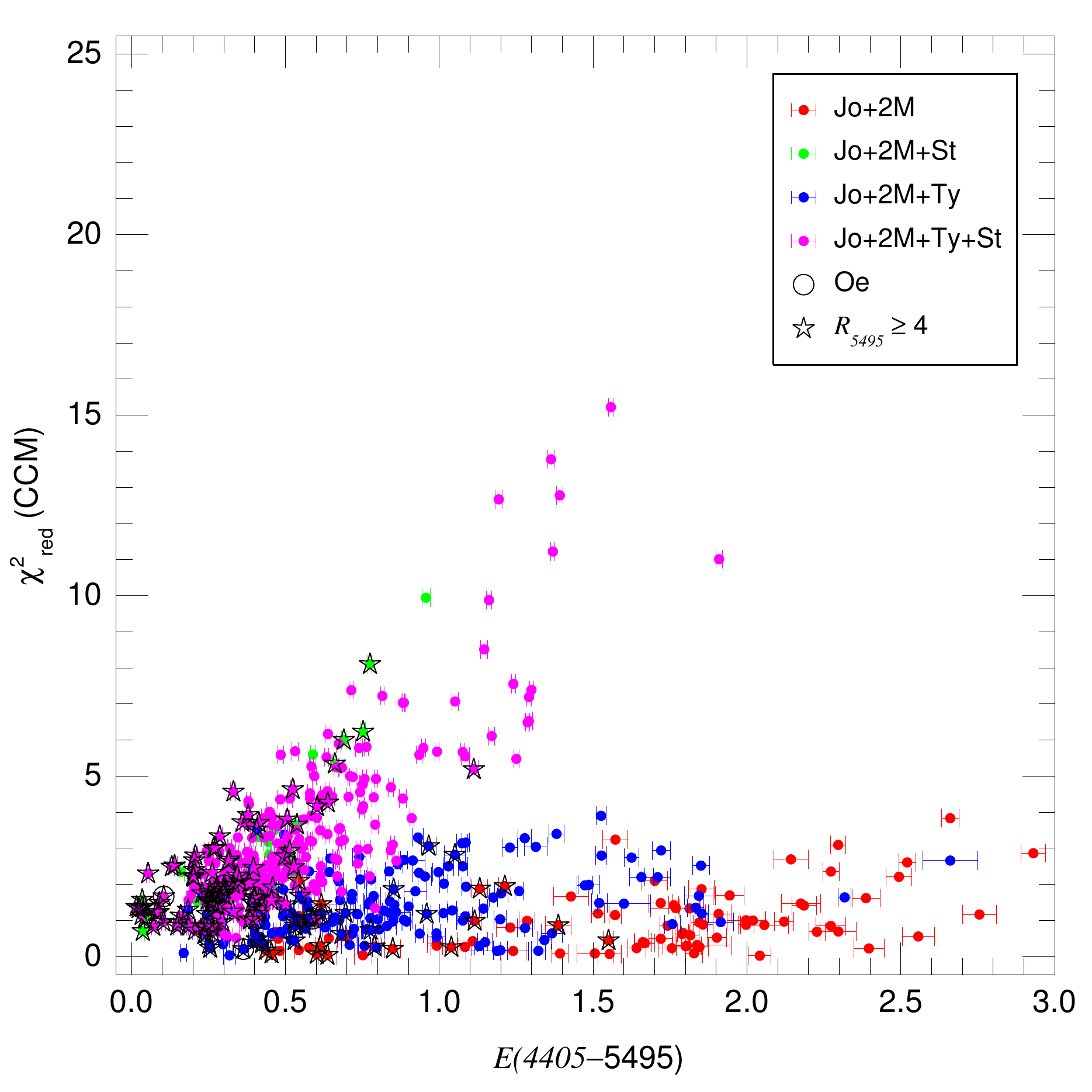}}
\centerline{\includegraphics*[width=0.55\linewidth]{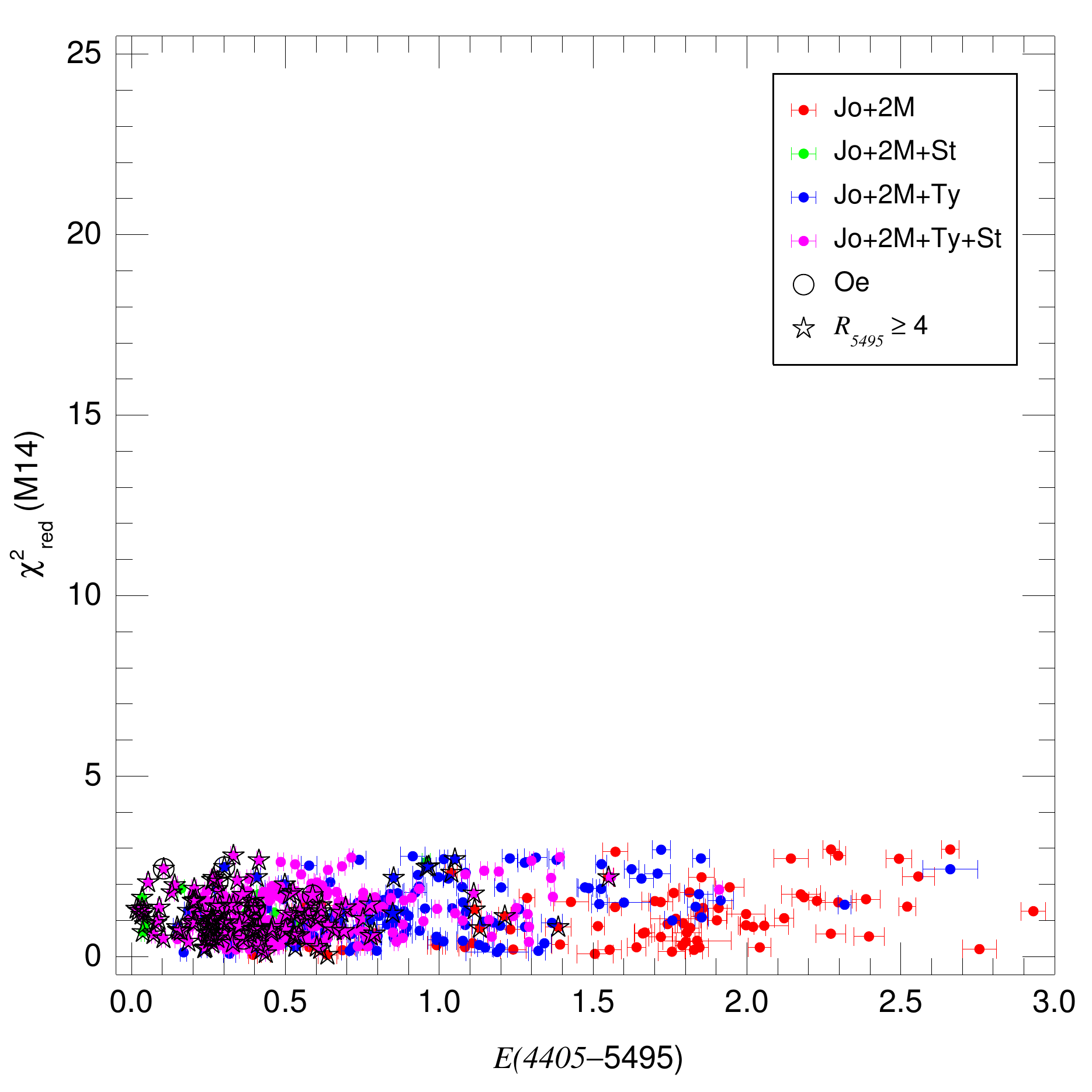}}
\caption{Reduced $\chi^2$ of the CHORIZOS fit to the photometry of the GOSSS I+II+III sample as a function of the amount of 
extinction $E(4405-5495)$. The top left, top right, and bottom panels are for the F99, CCM, and M14 families of extinction laws, respectively. Different 
colors are used to indicate the photometric bands available for each star (Jo = Johnson $UBV$, 2M = 2MASS $JHK_{\rm s}$, Ty = Tycho-2 $BV$, 
St = Str\"omgren $uvby$). Additional symbols are overplotted for Oe stars and objects with large values of $R_{5495}$ (i.e. extinction caused by large grain 
sizes).}
\label{figure1}
\end{figure}

\begin{figure}
\centerline{\includegraphics*[width=0.55\linewidth]{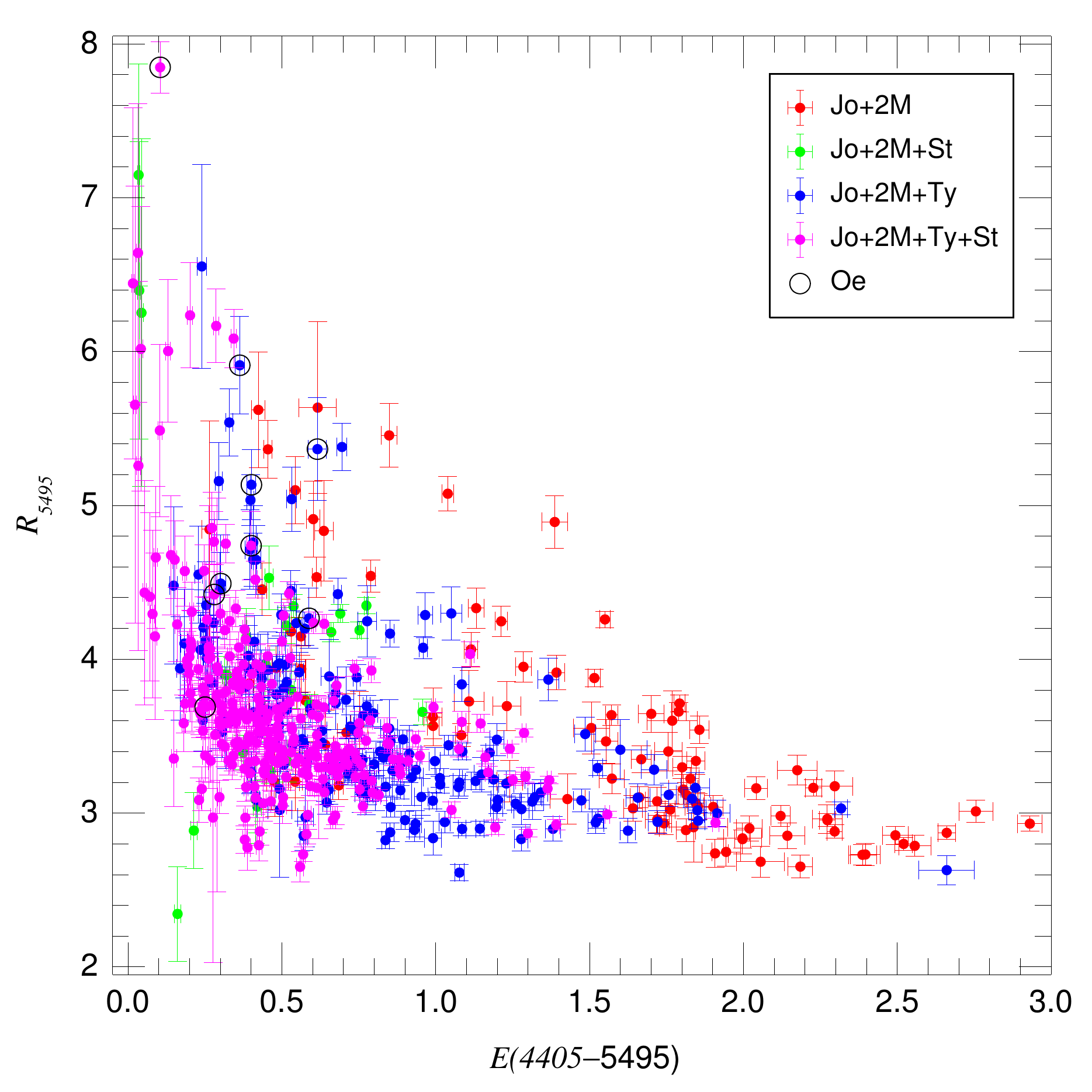}\,
            \includegraphics*[width=0.55\linewidth]{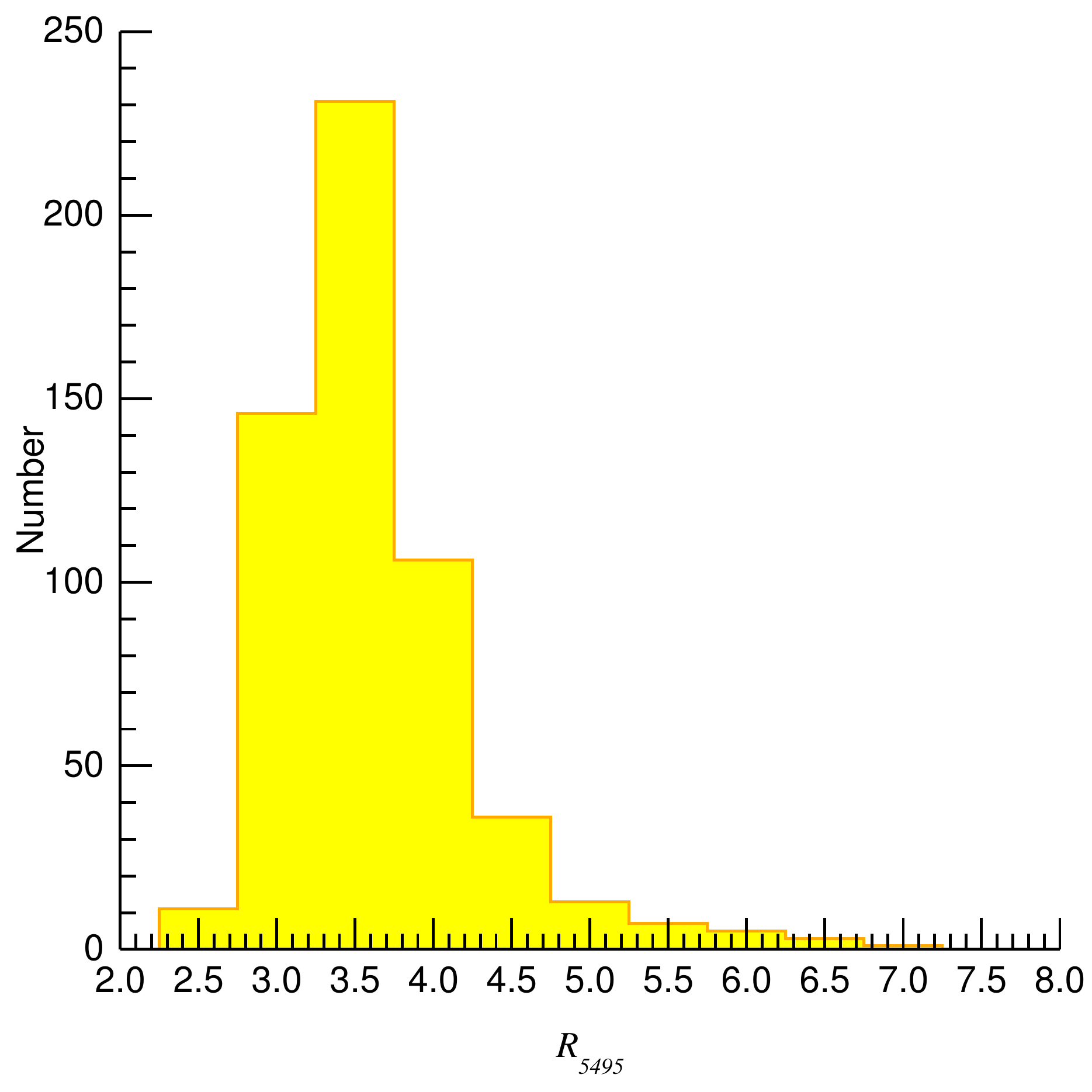}} 
\centerline{\includegraphics*[width=0.55\linewidth]{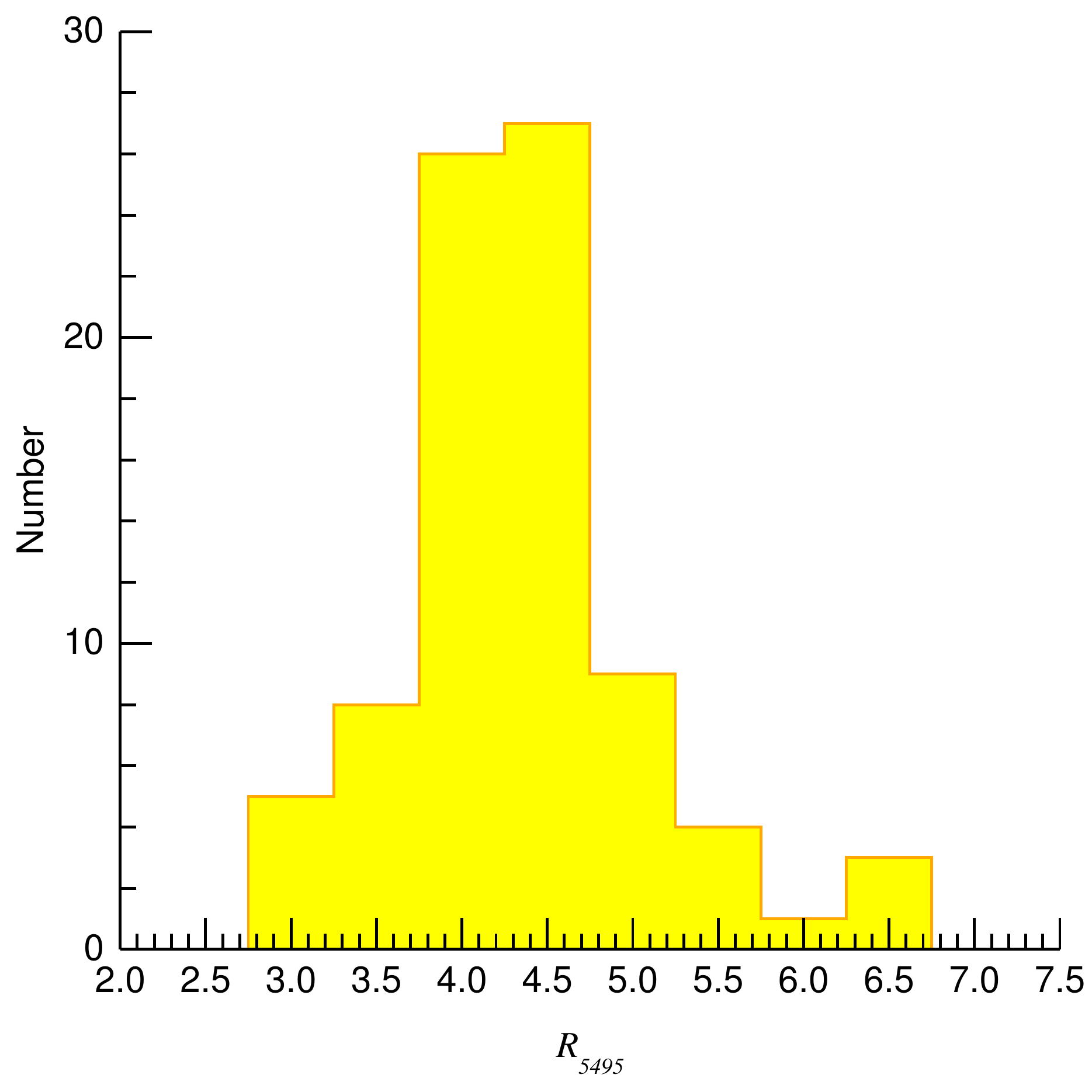}}
\caption{[top left] $R_{5495}$ as a function of $E(4405-5495)$ for the CHORIZOS fits for the GOSSS I+II+III sample using the M14 family of extinction 
laws.  See Fig.~\ref{figure1} for the symbol nomenclature. [top right] $R_{5495}$ histogram for the Galactic GOSSS I+II+III sample. [bottom] $R_{5495}$ 
histogram for the 30 Doradus VFTS sample of \href{http://adsabs.harvard.edu/abs/2014A&A...564A..63M}{Ma{\'\i}z Apell\'aniz et al. (2014a)}.}
\label{figure2}
\end{figure}

\section{Comparing extinction-law families}

$\,\!$\indent {\it Data.} We have collected optical-NIR photometry for the GOSSS I+II+III sample 
(\href{http://adsabs.harvard.edu/abs/2011ApJS..193...24S}{Sota et al. 2011}, \href{http://adsabs.harvard.edu/abs/2014ApJS..211...10S}{2014}, 
\href{http://adsabs.harvard.edu/abs/2016ApJS..224....4M}{Ma{\'\i}z Apell\'aniz et al. 2016}). All targets have Johnson $UBV$ and 2MASS $JHK_{\rm s}$, 
some have Tycho-2 $BV$ and/or Str\"omgren $uvby$.

{\it What have we done?} We fit the amount [$E(4405-5495)$] and type [$R_{5495}$] of extinction for each star using CHORIZOS 
(\href{http://adsabs.harvard.edu/abs/2004PASP..116..859M}{Ma{\'\i}z Apell\'aniz 2004}, \href{http://adsabs.harvard.edu/abs/2013hsa7.conf..657M}{2013b})
and three families of extinction laws: F99 (\href{http://adsabs.harvard.edu/abs/1999PASP..111...63F}{Fitzpatrick 1999}), CCM 
(\href{http://adsabs.harvard.edu/abs/1989ApJ...345..245C}{Cardelli et al. 1989}) and M14 
(\href{http://adsabs.harvard.edu/abs/2014A&A...564A..63M}{Ma{\'\i}z Apell\'aniz et al. 2014a}). See Fig~\ref{figure1}.

{\it Results.} We first study which family provides better fits to the data.

\begin{itemize}
 \item F99: poorest fits of all three families, especially for large values of $R_{5495}$.
 \item CCM: overall better than F99 except for targets with Str\"omgren photometry due to its use of a seventh-degree polynomial in $1/\lambda$
       for interpolation in the optical.
 \item M14: best fits of all, with a reduced $\chi^2$ always under 3.0.
\end{itemize}

We point out that even though the M14 family was derived using 30 Doradus data, it provides a better description of Galactic optical-NIR extinction than 
either the F99 or CCM families.

Once we have determined that the M14 family provides the best results of the three, we show in Fig.~\ref{figure2} some of the results found for
$E(4405-5495)$ and $R_{5495}$.

\begin{itemize}
 \item As previously known, the majority of Galactic sightlines have $R_{5495}$ between 3.0 and 3.5.
 \item At very low extinctions [$E(4405-5495) < 0.2$] the error bars on $R_{5495}$ are too large to yield significant results.
 \item There are few stars with $R_{5495} < 3.0$ (sightlines with many small dust grains) but they are dominant among objects with large extinction.
 \item In the range $0.2 < E(4405-5495) < 1.2$ there is a significant fraction of stars with large values of $R_{5495}$. Those sightlines are depleted in 
       small grains and are associated with H\,{\sc ii} regions.
 \item The $R_{5495}$ histograms for the Galaxy and 30 Doradus are markedly different, with the latter showing a larger fraction of high-$R_{5495}$ 
       sightlines.  The differences can be explained by [a] the lower values of $E(4405-5495)$ and [b] the larger fraction of H\,{\sc ii} region sightlines 
       in 30 Doradus.
\end{itemize}

\begin{figure}
\centerline{\includegraphics*[width=1.1\linewidth]{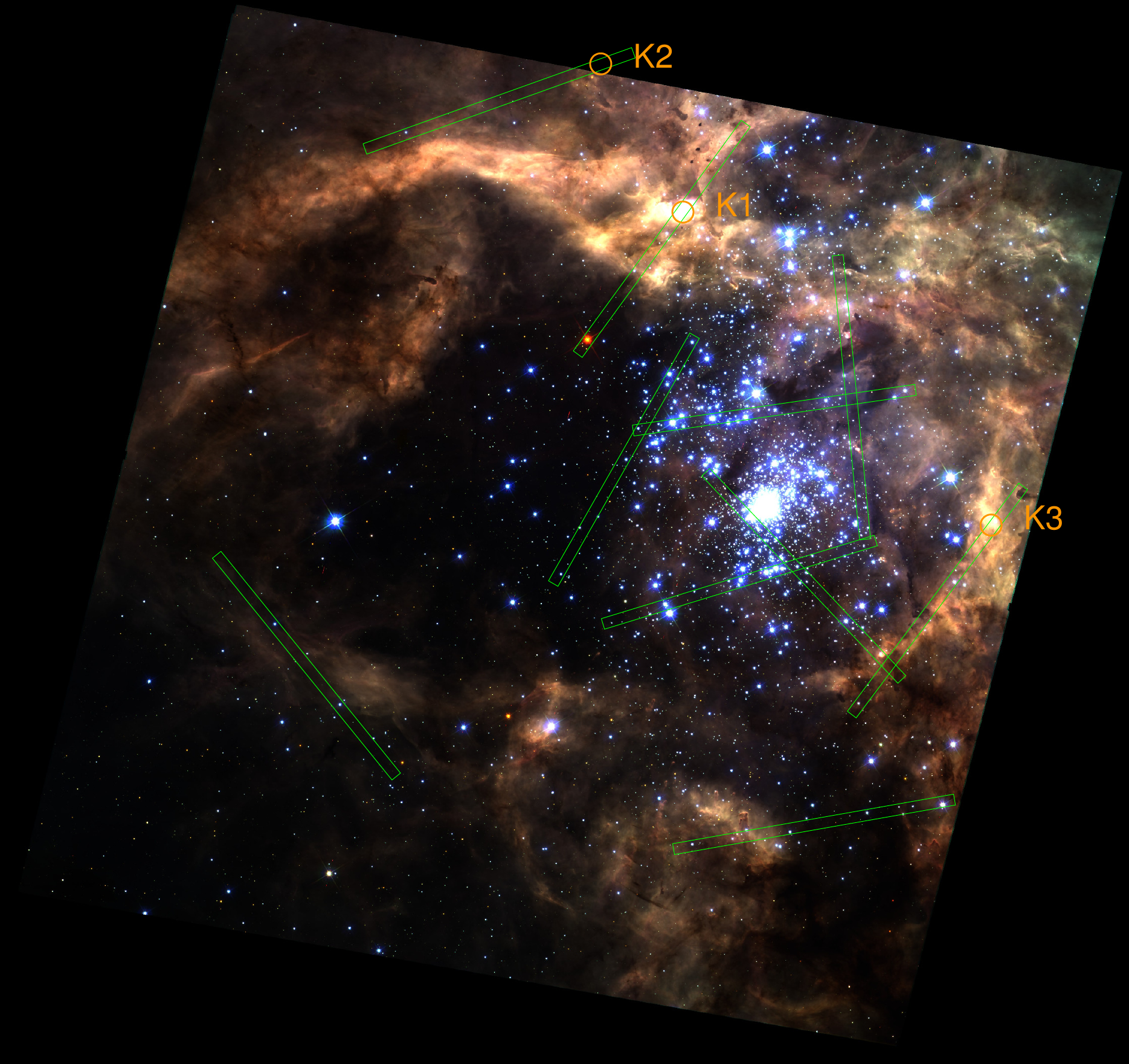}}
\caption{UVIS-WFC3 image of 30 Doradus with the ten STIS slits overimposed. The slit locations and PAs were selected to include a large variety of 
values of $R_{5495}$ and of environments, from stellar cocoons to the diffuse ISM. Knots 1, 2, and 3 from 
\href{http://adsabs.harvard.edu/abs/2002AJ....124.1601W}{Walborn et al. (2002)} are marked. The field size is 206\arcsec$\times$194\arcsec, N is top, 
and E is left.}
\label{figure3}
\end{figure}

\begin{figure}
\centerline{\includegraphics*[width=1.1\linewidth]{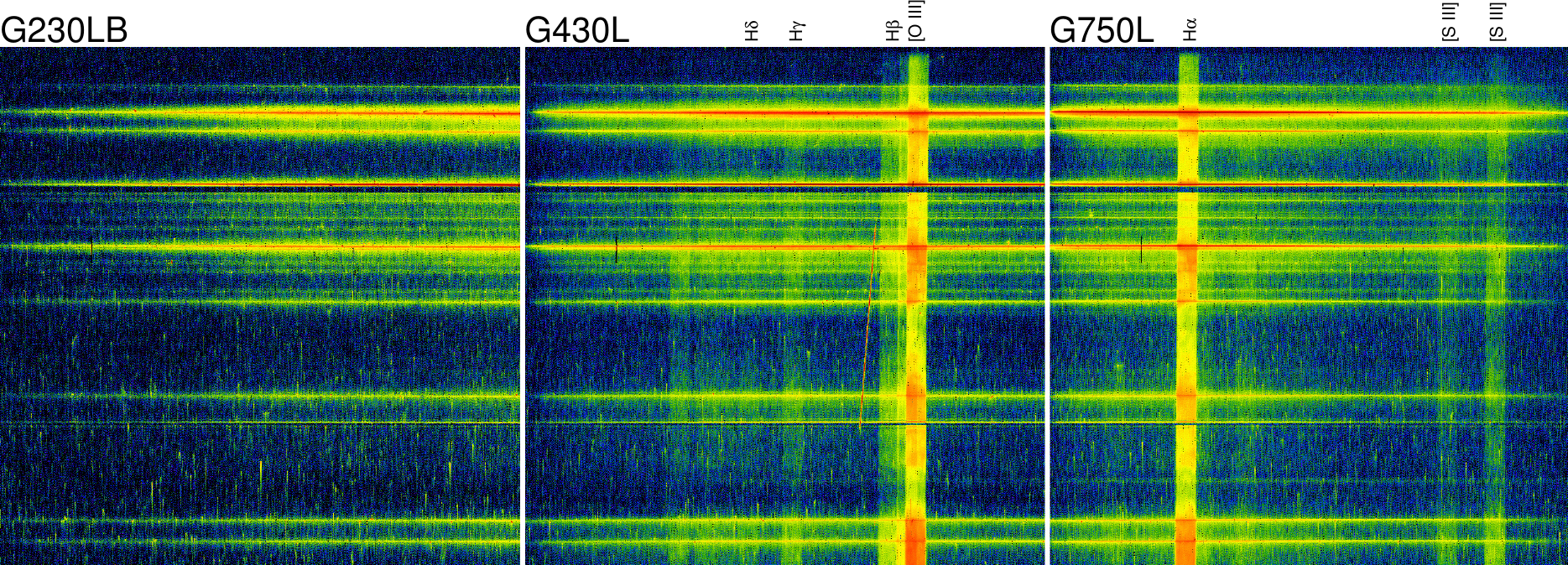}}
\caption{G230LB+G430L+G750L (1700-10\,200~\AA) STIS exposures of one of the ten 30 Doradus slits. The horizontal coordinate is wavelength and the
vertical one is slit position (covering 51\arcsec). This slit covers a region close to R136 which is relatively dense in OB stars and has an intermediate 
nebular intensity. The positions of some prominent nebular lines are marked. Note that since we are using a 2\arcsec-wide slit, each nebular line creates a 
rectangular image in the CCD. The two horizontal dark regions are caused by occulting bars.}
\label{figure4}
\end{figure}

\begin{figure}
\centerline{\includegraphics*[width=1.1\linewidth]{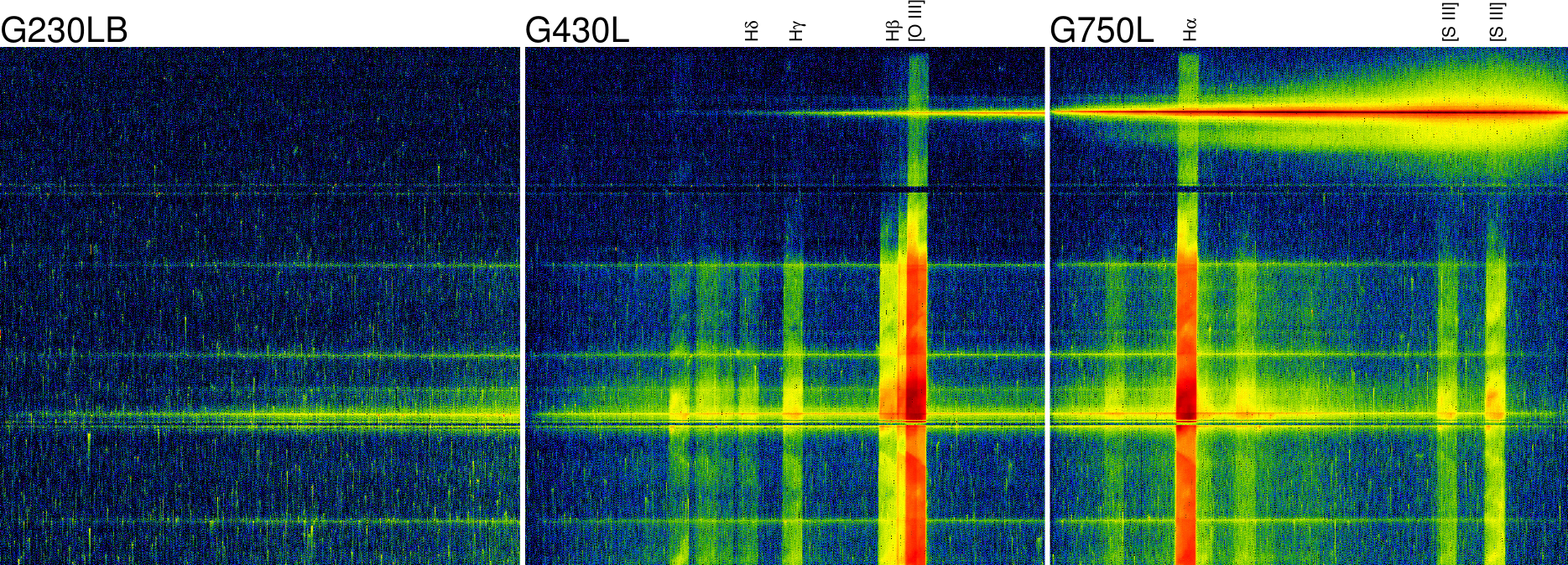}}
\caption{Same as Fig.~\ref{figure4} but for a different slit. The lower half has several OB stars immersed in intense nebulosity while the upper half is
dominated by a red supergiant in a region with little nebulosity.}
\label{figure5}
\end{figure}

\section{30 Doradus with STIS}

$\,\!$\indent We are obtaining 10 visits of HST/STIS spectrophotometry under GO program 14\,104 (9 of them have been executed as of October 2016, Fig.~\ref{figure3}).
For each visit we use three gratings, G230LB+G430L+G750L, in order to continuously cover the 1700-10\,200~\AA\ range. We use the 2\arcsec-wide slit in order
to obtain data for for 30 OB stars (Figs.~\ref{figure4}~and~\ref{figure5}). The stellar spectra are extracted with MULTISPEC 
(\href{http://adsabs.harvard.edu/abs/2005stis.rept....2M}{Ma{\'\i}z Apell\'aniz 2005b}) to correctly eliminate nebular contamination. The stars were selected
to probe different environments: H\,{\sc ii} region, diffuse ISM, and molecular gas. The questions we want to answer are:

\begin{itemize}
 \item What are the UV-NIR detailed ($\sim$100 \AA\ resolution) extinction laws? 
 \item Is the \href{http://adsabs.harvard.edu/abs/1958AJ.....63..201W}{Whitford (1958)} knee real and does it exist for all values of $R_{5495}$?
 \item Are there real differences between the Galactic and 30 Doradus UV extinction laws?
\end{itemize}

\section{Infrared extinction}

\begin{figure}
\centerline{\includegraphics*[width=0.55\linewidth]{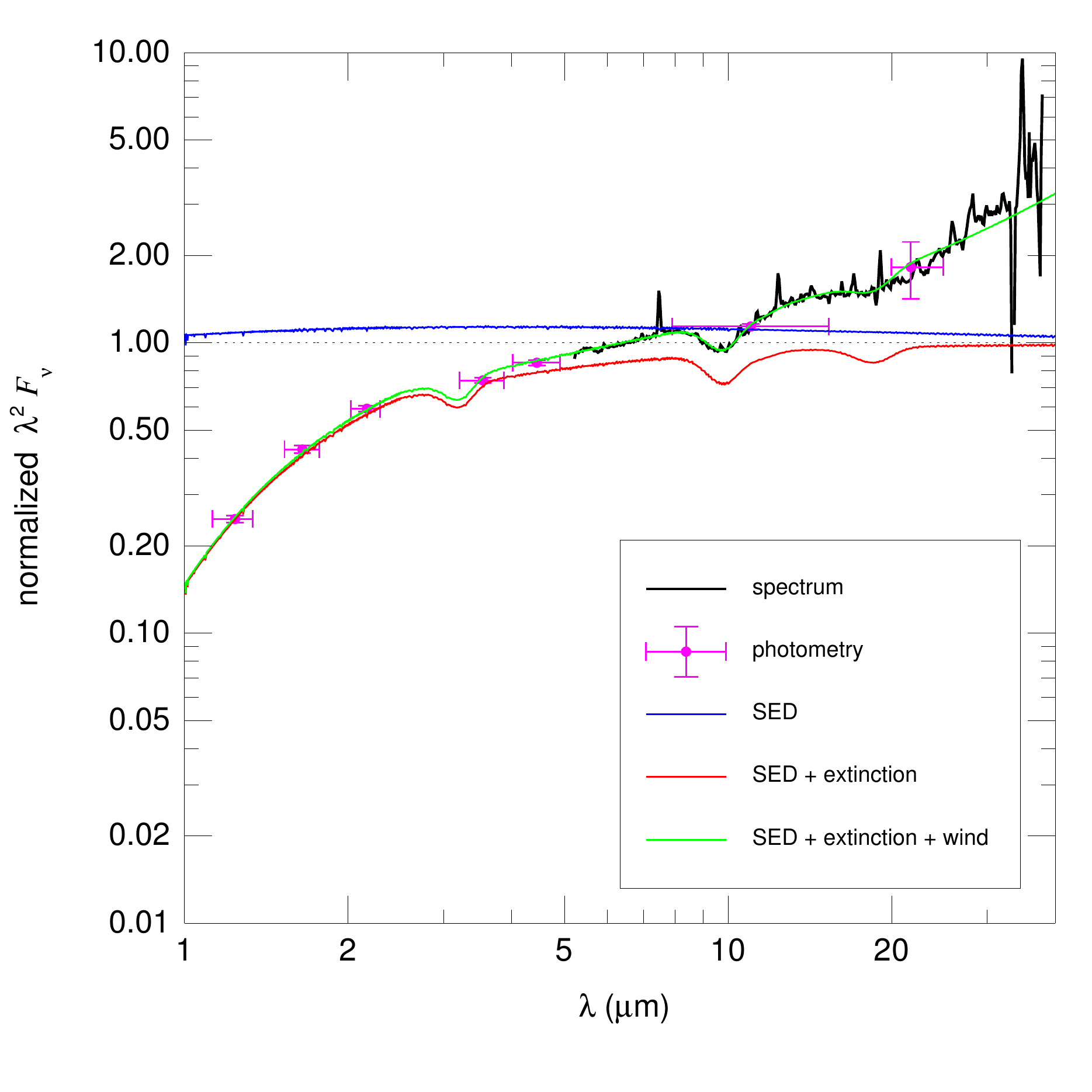} \,
            \includegraphics*[width=0.55\linewidth]{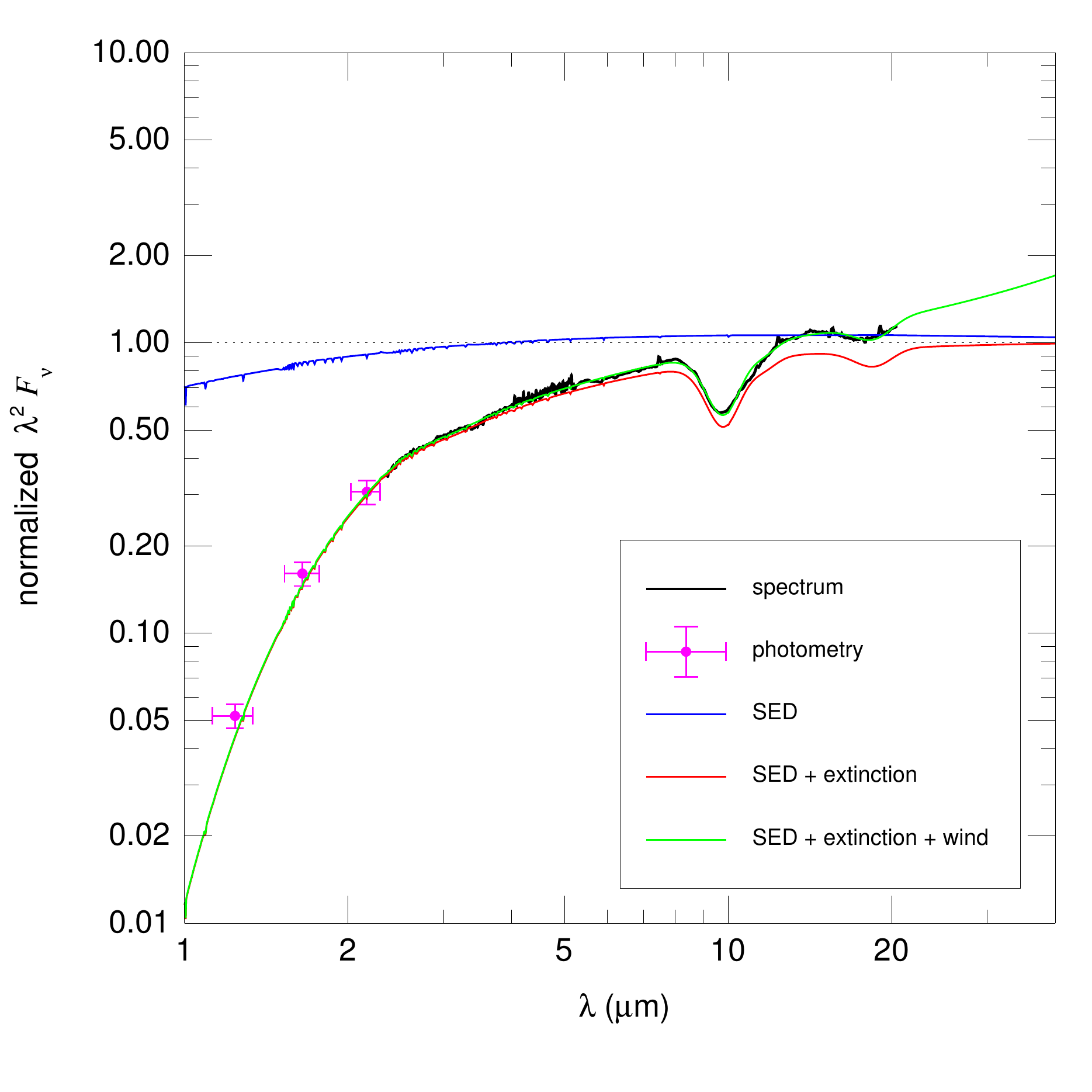}}
\caption{Two examples of atmosphere SED + wind + extinction law fits. [left] An O star with moderate extinction and
strong wind with 2MASS $JHK_{\rm s}$ and WISE W1+W2+W3+W4 photometry and Spitzer spectrophotometry. [right] A B star with high
extinction and moderate wind with 2MASS $JHK_{\rm s}$ photometry and ISO spectrophotometry. The vertical scale is $\lambda^2 F_\nu$ normalized to the 
atmosphere SED value at infinity.}
\label{figure6}
\end{figure}

$\,\!$\indent We aim to derive a detailed IR extinction law for OB stars in the solar neighborhood; tie it up with the optical extinction law finding 
appropriate sightlines; and analyze possible variations, checking if there is a relationship with $R_{5495}$. Our strategy is:

\begin{itemize} 
 \item Collect Spitzer+ISO spectrophotometry and Spitzer photometry for GOSSS stars.
 \item Collect and reprocess WISE photometry.
 \item Use WISE imaging to determine the type of environment.
 \item Combine the data to fit: [a] atmosphere SED, [b] wind, and [c] extinction law for each sightline (Fig.~\ref{figure6}).
\end{itemize}

In the process of measuring the MIR extinction law, we collected and reprocessed the WISE photometry for the GOSSS I+II+III sample.
We tested the validity of the W1+W2+W3+W4 photometry comparing the results from the All-Sky 
(\href{http://adsabs.harvard.edu/abs/2012yCat.2307....0C}{Cutri et al. 2012}) and AllWISE 
(\href{http://adsabs.harvard.edu/abs/2013yCat.2328....0C}{Cutri et al. 2013}) data releases.
We compared the PSF photometry from the data releases with new aperture photometry using a realistic sky evaluation.
We also derived saturation corrections and identified cases with large uncertainties.
We will use our results to study the relationship between extinction and the environment.

\begin{figure}
\centerline{\includegraphics*[width=1.1\linewidth]{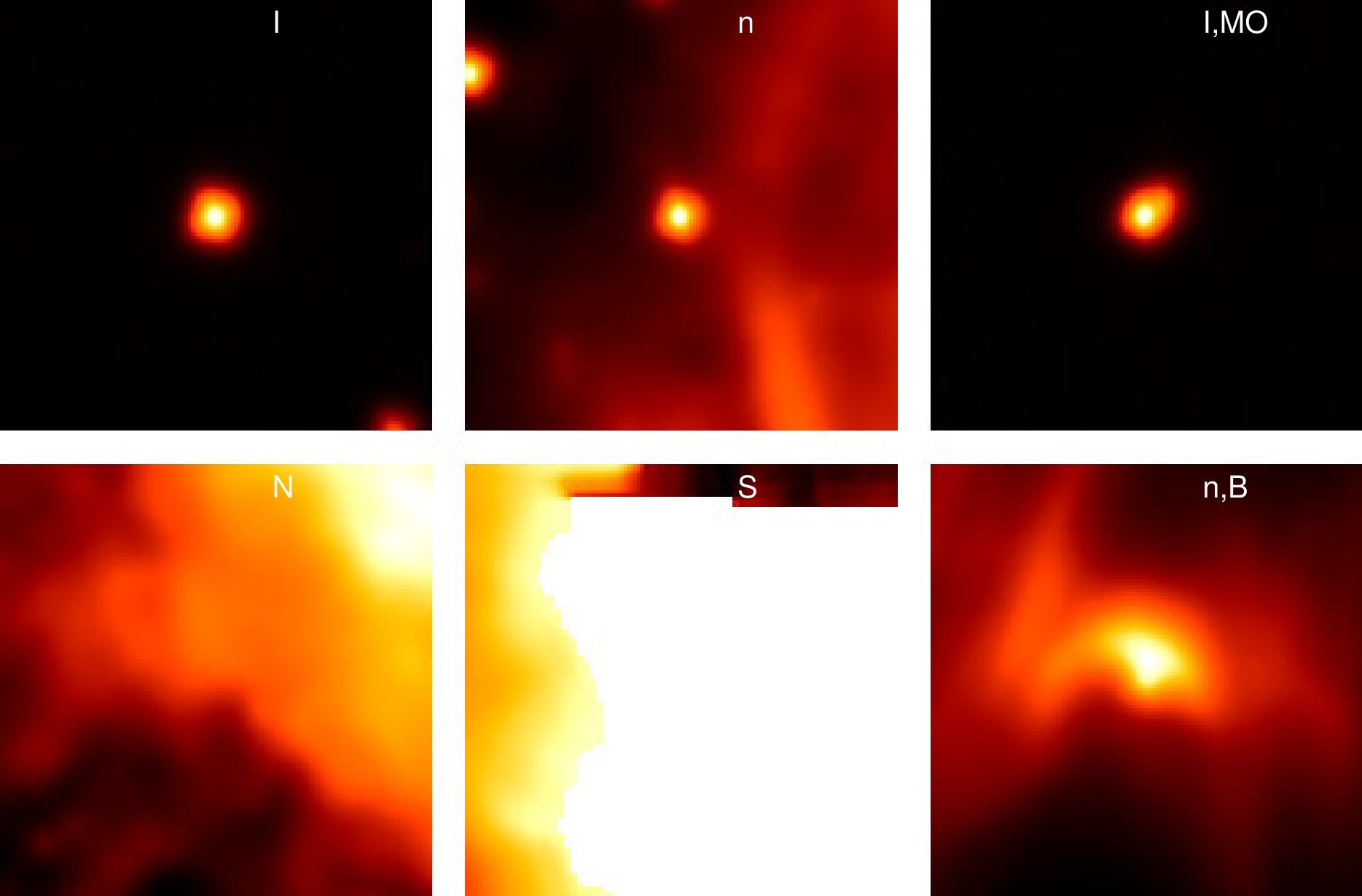}}
\caption{WISE W3 images and morphological classifications for six GOSSS I+II+III stars. Each field is 3\arcmin$\times$3\arcmin, with N towards the top
and E towards the left, the target at the center, and a logarithmic intensity scale.}
\label{figure7}
\end{figure}

\begin{itemize} 
 \item WISE W3+W4 can be used to identify the presence of the warm dust associated with H\,{\sc ii} regions, detecting 
       them even at high extinctions.
 \item We morphologically classified the sample using the WISE images (Fig.~\ref{figure7}) in order to study
       possible variations in the extinction law.
 \item We use the following main flags: I (isolated), n (weak nebulosity), N (strong nebulosity), S (saturated background i.e. bright H\,{\sc ii} region).
 \item We use the following optional flags: MO (hidden multiplicity in WISE, optical companion), MI (hidden multiplicity in WISE, IR companion), B (bow shock visible).
\end{itemize}

\section*{References}

\begin{itemize}
 \item \href{http://adsabs.harvard.edu/abs/2006MNRAS.366..739A}{Arias, J. I. et al. 2006, {\it MNRAS} {\bf 366}, 739}.
 \item \href{http://adsabs.harvard.edu/abs/1989ApJ...345..245C}{Cardelli, J. A. et al. 1989, {\it ApJ} {\bf 345}, 245}.
 \item \href{http://adsabs.harvard.edu/abs/2012yCat.2307....0C}{Cutri, R. M. et al. 2012, {\it VizieR Online Catalog} II/311}.
 \item \href{http://adsabs.harvard.edu/abs/2013yCat.2328....0C}{Cutri, R. M. et al. 2013, {\it VizieR Online Catalog} II/328}.
 \item \href{http://adsabs.harvard.edu/abs/1999PASP..111...63F}{Fitzpatrick, E. L. 1999, {\it PASP} {\bf 111}, 63}.
 \item \href{http://adsabs.harvard.edu/abs/2004PASP..116..859M}{Ma{\'{\i}}z Apell{\'a}niz, J. 2004, {\it PASP} {\bf 116}, 859}.
 \item \href{http://adsabs.harvard.edu/abs/2005PASP..117..615M}{Ma{\'{\i}}z Apell{\'a}niz, J. 2005a, {\it PASP} {\bf 117}, 615}.
 \item \href{http://adsabs.harvard.edu/abs/2005stis.rept....2M}{Ma{\'{\i}}z Apell{\'a}niz, J. 2005b, {\it STIS-ISR} 2005-02}.
 \item \href{http://adsabs.harvard.edu/abs/2006AJ....131.1184M}{Ma{\'{\i}}z Apell{\'a}niz, J. 2006, {\it AJ} {\bf 131}, 1184}.
 \item \href{http://adsabs.harvard.edu/abs/2007ASPC..364..227M}{Ma{\'{\i}}z Apell{\'a}niz, J. 2007, {\it ASPC} {\bf 364}, 227}.
 \item \href{http://adsabs.harvard.edu/abs/2012A&A...541A..54M}{Ma{\'{\i}}z Apell{\'a}niz, J. \& Rubio, M. 2012, {\it A\&A} {\bf 541}, 54}.
 \item \href{http://adsabs.harvard.edu/abs/2013hsa7.conf..583M}{Ma{\'{\i}}z Apell{\'a}niz, J. 2013a, {\it Highlights of Spanish Astrophysics VII}, 583}. 
 \item \href{http://adsabs.harvard.edu/abs/2013hsa7.conf..657M}{Ma{\'{\i}}z Apell{\'a}niz, J. 2013b, {\it Highlights of Spanish Astrophysics VII}, 657}. 
 \item \href{http://adsabs.harvard.edu/abs/2014A&A...564A..63M}{Ma{\'{\i}}z Apell{\'a}niz, J. et al. 2014a, {\it A\&A} {\bf 564}, 63}.
 \item \href{http://adsabs.harvard.edu/abs/2014IAUS..297..117M}{Ma{\'{\i}}z Apell{\'a}niz, J. et al. 2014b, {\it IAUS} {\bf 297}, 117}.
 \item \href{http://adsabs.harvard.edu/abs/2015hsa8.conf..402M}{Ma{\'{\i}}z Apell{\'a}niz, J. 2015a, {\it Highlights of Spanish Astrophysics VIII}, 402}. 
 \item \href{http://adsabs.harvard.edu/abs/2015MmSAI..86..553M}{Ma{\'{\i}}z Apell{\'a}niz, J. 2015b, {\it MmSAI} {\bf 86}, 553}.
 \item \href{http://adsabs.harvard.edu/abs/2015A&A...579A.108M}{Ma{\'{\i}}z Apell{\'a}niz, J. et al. 2015a, {\it A\&A} {\bf 579}, 108}.
 \item \href{http://adsabs.harvard.edu/abs/2015A&A...583A.132M}{Ma{\'{\i}}z Apell{\'a}niz, J. et al. 2015b, {\it A\&A} {\bf 583}, 132}.
 \item \href{http://adsabs.harvard.edu/abs/2015hsa8.conf..604M}{Ma{\'{\i}}z Apell{\'a}niz, J. et al. 2015c, {\it Highlights of Spanish Astrophysics VIII}, 604}. 
 \item \href{http://adsabs.harvard.edu/abs/2016ApJS..224....4M}{Ma{\'{\i}}z Apell{\'a}niz, J. et al. 2016, {\it ApJS} {\bf 224}, 4}. 
 \item \href{http://adsabs.harvard.edu/abs/2011ApJS..193...24S}{Sota, A. et al. 2011, {\it ApJS} {\bf 193}, 24}. 
 \item \href{http://adsabs.harvard.edu/abs/2014ApJS..211...10S}{Sota, A. et al. 2014, {\it ApJS} {\bf 211}, 10}. 
 \item \href{http://adsabs.harvard.edu/abs/2002AJ....124.1601W}{Walborn, N. R. et al. 2002, {\it AJ} {\bf 124}, 1601}.
 \item \href{http://adsabs.harvard.edu/abs/1958AJ.....63..201W}{Whitford, A. E. 1958, {\it AJ} {\bf 63}, 201}.
\end{itemize}

\end{document}